\documentclass[floats, prd, eqnum, showpacs, nofootinbib, twocolumn,eqsecnum]{revtex4-1}

\usepackage{color,graphicx}
\usepackage{amsfonts}
\usepackage{amssymb}
\usepackage{comment}
\begin{document}

\title{Gravitational lensing by a photon sphere in a Reissner-Nordstr\"{o}m naked singularity spacetime in strong deflection limits} 
\author{Naoki Tsukamoto${}^{1}$}\email{tsukamoto@rikkyo.ac.jp}

\affiliation{
${}^{1}$Department of General Science and Education, National Institute of Technology, Hachinohe College, Aomori 039-1192, Japan \\
}

\begin{abstract}
We investigate gravitational lensing by a photon sphere in a Reissner-Nordstr\"{o}m naked singularity spacetime in strong deflection limits.
Because of the nonexistence of an event horizon and the existence of a potential barrier near an antiphoton sphere,
infinite numbers of images slightly inside and outside of a photon sphere can appear.
We obtain the analytic expressions of the factors of logarithmic divergent terms and the constant terms of the deflection angles 
in the strong deflection limits not only the for the little outside but also the barely inside of the photon sphere
without Taylor expansions in the power of an electric charge. 
We can distinguish between a Reissner-Nordstr\"{o}m black hole spacetime and the naked singularity spacetime since
the images little inside of the photon sphere around the naked singularity 
are significantly brighter than the image barely outside of the photon sphere 
around the black hole and the naked singularity.
\end{abstract}
\maketitle

\section{Introduction}
Recently, gravitational waves emitted by black holes have been detected directly
 by LIGO Scientific and Virgo Collaborations~\cite{Abbott:2016blz}
and a dark shadow image in a bright gas around a supermassive black hole candidate at center of a giant elliptical galaxy M87 
has been reported by Event Horizon Telescope Collaboration~\cite{Akiyama:2019cqa}. 
Because of the recent observations, to understand phenomena in a strong gravitational field described by general relativity will be more important than before.

Black holes and the other compact objects which can be black hole mimickers 
have unstable (stable) light circular orbits called photon sphere (antiphoton sphere)~\cite{Perlick_2004_Living_Rev,Claudel:2000yi,Perlick:2021aok,
Hod:2017xkz,Sanchez:1977si,Decanini:2010fz,Wei:2011zw,Press:1971wr,Goebel_1972,Stefanov:2010xz,Raffaelli:2014ola,Abramowicz_Prasanna_1990,Abramowicz:1990cb,Allen:1990ci,Hasse_Perlick_2002,Mach:2013gia,Chaverra:2015bya,Cvetic:2016bxi,Koga:2016jjq,Koga:2018ybs,Koga:2019teu,Barcelo:2000ta,Koga:2020gqd,Ames_1968,Synge:1966okc,Yoshino:2019qsh} 
or its alternatives and generalized surfaces~\cite{Claudel:2000yi,Gibbons:2016isj,Cunha:2017eoe,Shiromizu:2017ego,Yoshino:2017gqv,Galtsov:2019bty,Galtsov:2019fzq,Koga:2019uqd,Siino:2019vxh,Yoshino:2019dty,Cao:2019vlu,Yoshino:2019mqw,Lee:2020pre,Izumi:2021hlx,Siino:2021kep}
because of their strong gravitation.
It is important to focus on the photon sphere and the antiphoton sphere since
the circular light orbits have close ties with both observational and theoretical aspects and 
stable light circular orbits of ultracompact objects may cause instability because of the slow decay of linear waves~\cite{Keir:2014oka,Cardoso:2014sna,Cunha:2017qtt}.

Gravitational lensing~\cite{Schneider_Ehlers_Falco_1992,Schneider_Kochanek_Wambsganss_2006}, which is a phenomenon that 
light rays are bended by a lensing object in front of a source object, can be used to survey 
dark and massive objects not only in weak gravitational fields but also in strong gravitational fields.
Images of light rays scattered by a photon sphere in a Schwarzschild spacetime 
were studied by Hagihara in 1931~\cite{Hagihara_1931} 
and by Darwin in 1959~\cite{Darwin_1959} independently 
and then images reflected by the photon sphere were revived many times 
\cite{Atkinson_1965,Luminet_1979,Ohanian_1987,Nemiroff_1993,Virbhadra:1998dy,Frittelli_Kling_Newman_2000,Virbhadra_Ellis_2000,Bozza_Capozziello_Iovane_Scarpetta_2001,Bozza:2002zj,Virbhadra:2002ju,
Perlick:2003vg,Virbhadra:2008ws,Bozza_2010,Tsukamoto:2016zdu,Shaikh:2019jfr,Shaikh:2019itn,Tsukamoto:2020uay,Tsukamoto:2020iez,Paul:2020ufc}.

The deflection angle $\alpha$ of a light in a strong deflection limit $b\rightarrow b_\mathrm{m}+0$ 
in a general asymptotically flat, static, and spherical symmetric spacetime
has a form, as shown by Bozza~\cite{Bozza:2002zj}, 
\begin{eqnarray}\label{eq:1}
\alpha\left(b\right) &=&- \bar{a} \log \left( \frac{b}{b_\mathrm{m}}-1 \right) +\bar{b} \nonumber\\
&&+O \left( \left( \frac{b}{b_\mathrm{m}}-1 \right) \log \left( \frac{b}{b_\mathrm{m}}-1 \right) \right),
\end{eqnarray}
where $b$ is the impact parameter of the ray, $b_\mathrm{m}$ is a critical impact parameter, 
and $\bar{a}$ and $\bar{b}$ can be described by parameters of the spacetime.\footnote{ %
The order of error terms estimated as $O\left(b-b_\mathrm{m}\right)$ in Ref.~\cite{Bozza:2002zj}
should be read as $O \left( \left( b/b_\mathrm{m}-1 \right) \log \left( b/b_\mathrm{m}-1 \right) \right)$ 
as discussed in Refs.~\cite{Iyer:2006cn,Tsukamoto:2016qro,Tsukamoto:2016jzh}. 
We can see the explicit form of the Schwarzschild spacetime~\cite{Iyer:2006cn}. 
}~
The strong-deflection-limit analysis has been applied to many black hole, wormhole, and naked singularity spacetimes 
with the photon spheres and the analysis has been extended
and alternative analysis have been suggested~\cite{Tsukamoto:2016zdu,Shaikh:2019jfr,Shaikh:2019itn,Tsukamoto:2020uay,
Tsukamoto:2020iez,Paul:2020ufc,Bozza:2002af,Eiroa:2002mk,Petters:2002fa,Eiroa:2003jf,Bozza:2004kq,Bozza:2005tg,Bozza:2006sn,Bozza:2006nm,Iyer:2006cn,
Bozza:2007gt,Tsukamoto:2016qro,Ishihara:2016sfv,Tsukamoto:2016oca,Tsukamoto:2016jzh,Tsukamoto:2017edq,Hsieh:2021scb,Aldi:2016ntn,
Tsukamoto:2020bjm,Takizawa:2021gdp,Tsukamoto:2021caq,Aratore:2021usi}.

We emphasize that it is important to find exact expressions for $\bar{a}$ and $\bar{b}$
since observables in the strong deflection limit 
are characterized by the parameters $\bar{a}$ and $\bar{b}$ 
and they might give us a hint to understand relations 
between the gravitational lensing and other phenomena in the strong gravitational fields.
The coefficient $\bar{a}$ is often obtained analytically in the analysis by Bozza~\cite{Bozza:2002zj} while 
a part of the term $\bar{b}$ usually is calculated in numerical or
calculated analytically after an expansion by a parameter of a spacetime with a few exceptions:
Bozza~\textit{et al.}~\cite{Bozza_Capozziello_Iovane_Scarpetta_2001}
and Bozza~\cite{Bozza:2002zj} have obtained the exact form of $\bar{a}$ and $\bar{b}$ in the Schwarzschild spacetime.
And, in Ref.~\cite{Iyer:2006cn}, it has been shown that 
the deflection angle in Refs.~\cite{Bozza_Capozziello_Iovane_Scarpetta_2001,Bozza:2002zj} is equivalent to the one by Darwin~\cite{Darwin_1959}. 
The exact forms of $\bar{a}$ and $\bar{b}$ in a braneworld black hole spacetime have been obtained by Eiroa~\cite{Eiroa:2004gh},
exact ones in 5-dimensional and 7-dimensional Schwarzschild spacetime have been calculated by Tsukamoto~\textit{et al.}~\cite{Tsukamoto:2014dta}.
Tsukamoto~\cite{Tsukamoto:2016jzh,Tsukamoto:2016qro,Tsukamoto:2016zdu} has extended Bozza's method~\cite{Bozza:2002zj} for ultrastatic spacetimes, 
which has a constant norm of a time-translational Killing vector, 
and has obtained exact forms of $\bar{a}$ and $\bar{b}$ in an Ellis wormhole spacetime 
without an Arnowitt-Deser-Misner~(ADM) mass~\cite{Tsukamoto:2016jzh,Tsukamoto:2016qro}~\footnote{ %
Bhattacharya and Potapov have considered a deflection angle in the strong deflection limit in an Ellis-Bronnikov wormhole spacetime 
with an ADM mass by using Bozza's method
and then obtained the same deflection angle in the strong deflection limit as Refs.~\cite{Tsukamoto:2016jzh,Tsukamoto:2016qro}
 as a massless case~\cite{Bhattacharya:2019kkb}.
}~%
and Tsukamoto and Harada have obtained exact ones in an ultrastatic wormhole spacetime~\cite{Tsukamoto:2016zdu}. 

Astrophysical objects in nature would not have the large amount of an electrical charge since they are quickly neutralized.
In general relativity, however, the Reissner-Nordstr\"{o}m spacetime is often considered as a simple toy model of a compact object 
since it has a richer structure than the Schwarzschild spacetime
and since it could be treated analytically as well as the non-charged case.
A shadow image~\cite{deVries:2000,Takahashi:2005hy}, the time delay of light rays~\cite{Sereno:2003nd}, 
gravitational lensing~\cite{Bin-Nun:2010exl,Bin-Nun:2010lws}, and retrolensing~\cite{Eiroa:2003jf,Tsukamoto:2016oca} by a Reissner-Nordstr\"{o}m black hole
have been investigated. 
Chiba and Kimura have investigated the deflection angle of a light in a Hayward metric and they have pointed out that the qualitative behavior of null geodesics in Hayward metric 
is almost the same as a behavior in the Reissner-Nordstr\"{o}m spacetime~\cite{Chiba:2017nml}. 
This implies the other compact objects with a charge also have a similar behavior.

Eiroa \textit{et al.} calculated the deflection angle in the strong deflection limit in numerical~\cite{Eiroa:2002mk} in the Reissner-Nordstr\"{o}m spacetime. 
In Ref.~\cite{Bozza:2002zj}, the term $\bar{b}$ was calculated partly in numerical 
and calculated analytically by expanding by the electrical charge 
by Bozza and $\bar{b}$ cannot be obtained as an exact form without the Taylor expansion on the charge.
Tsukamoto~\cite{Tsukamoto:2016jzh} and Tsukamoto and Gong~\cite{Tsukamoto:2016oca}
have suggested the alternative method of Bozza's method 
to obtain the exact form of $\bar{b}$. 
They have showed that the exact forms of $\bar{a}$ and $\bar{b}$ in analytical 
are equivalent with the numerical results by Eiroa \textit{et al.}~\cite{Eiroa:2002mk} and by Bozza~\cite{Bozza:2002zj}.
Bad\'\i{}a and Eiroa have obtained exact forms of $\bar{a}$ and $\bar{b}$ in a Horndeski black hole spacetime 
by using the alternative method~\cite{Badia:2017art}. 
Recently, exact forms of $\bar{a}$ and $\bar{b}$ in Kerr and Kerr-Newman spacetimes 
on an equatorial plane have been obtained by Hsieh~\textit{et al.}~\cite{Hsieh:2021scb}.

The Reissner-Nordstr\"{o}m spacetime for $1<q/m<3/(2\sqrt{2})$,  
where $q$ and $m$ are its electrical charge and its mass, respectively,
does not have an event horizon but it has an antiphoton sphere and a photon sphere.
The photon sphere, the shadow, and the magnifications of lensed images 
have been studied in Refs.~\cite{Zakharov:2014lqa,Shaikh:2019itn}.
For $q/m=3/(2\sqrt{2})$, the antiphoton sphere and the photon sphere degenerate to be a marginally unstable photon sphere 
and its gravitational lensing has been considered in Refs.~\cite{Chiba:2017nml,Tsukamoto:2020iez}.  

Shaikh~\textit{et al.} have considered gravitational lensing by compact objects with an antiphoton sphere and a photon sphere 
and without an event horizon 
in a strong deflection limit $b\rightarrow b_\mathrm{m}-0$ in a general asymptotically flat, static, and spherical symmetric spacetime~\cite{Shaikh:2019itn}. 
The deflection angle of a light ray which is reflected by a potential barrier near the antiphoton sphere 
in the strong deflection limit~$b\rightarrow b_\mathrm{m}-0$ has a form  
\begin{eqnarray}\label{eq:2}
\alpha\left(b\right) &=&- \bar{c} \log \left( \frac{b_\mathrm{m}^2}{b^2}-1 \right) +\bar{d} \nonumber\\
&&+O \left( \left( \frac{b_\mathrm{m}}{b}-1 \right) \log \left( \frac{b_\mathrm{m}}{b}-1 \right) \right),
\end{eqnarray}
where $\bar{c}$ and $\bar{d}$ can be characterized by the parameters of the spacetime if the photon sphere exists.~\footnote{%
We can approximate
\begin{eqnarray}
 \frac{b_\mathrm{m}^2}{b^2}-1 \sim 2 \left( \frac{b_\mathrm{m}}{b}-1 \right) \sim 2 \left( 1- \frac{b}{b_\mathrm{m}} \right). 
\end{eqnarray}
However, we use the form of Eq.(\ref{eq:2}) so that we make the error small as well as Ref.~\cite{Shaikh:2019itn}.}%
They have applied it to the Reissner-Nordstr\"{o}m naked singularity spacetime with $q^2/m^2=1.05$ 
but they do not show the explicit forms of $\bar{c}$ and $\bar{d}$.

On this paper, we investigate the details of the gravitational lensing in the strong deflection limits~$b\rightarrow b_\mathrm{m}-0$ and $b\rightarrow b_\mathrm{m}+0$
in the Reissner-Nordstr\"{o}m naked singularity spacetime with the antiphoton sphere and the photon sphere with $1<q/m<3/(2\sqrt{2})$
by using methods in Refs.~\cite{Bozza:2002zj,Tsukamoto:2016jzh,Shaikh:2019itn,Tsukamoto:2021caq}.
We obtain the exact forms of not only the factor $\bar{a}$ and the term $\bar{b}$ in Eq.~(\ref{eq:1}) 
but also the factor $\bar{c}$ and the term $\bar{d}$ in Eq.~(\ref{eq:2})
and we apply it to a supermassive black hole candidate at the center of our galaxy to calculate observables. 

This paper is organized as follows. 
We investigate the deflection angle in the Reissner-Nordstr\"{o}m spacetime in Sec.~II.
And we consider the one and observables in the strong deflection limits in Secs.~III and IV, respectively.
We give a conclusion in Sec V.
We review a weak-field approximation very shortly in appendix A.
We use the units in which the light speed and Newton's constant are unity.

\section{Deflection angle in the Reissner-Nordstr\"{o}m spacetime}
The line element of a Reissner-Nordstr\"{o}m spacetime is given by
\begin{equation}
\mathrm{d}s^2=-A(r)\mathrm{d}t^2+\frac{\mathrm{d}r^2}{A(r)}+r^2 \left(\mathrm{d}\vartheta^2+\sin^2 \vartheta \mathrm{d}\varphi^2\right),
\end{equation}
where $A(r)$ is given by
\begin{equation}
A(r)\equiv 1-\frac{2m}{r}+\frac{q^2}{r^2}
\end{equation}
and $m\geq 0$ is an ADM mass and $q\geq 0$ is an electrical charge.
It has an event horizon at $r=r_\mathrm{H}$, where
\begin{equation}
r_\mathrm{H}\equiv m+\sqrt{m^{2}-q^{2}},
\end{equation}
for $q\leq m$, 
and it has a naked singularity for $m<q$.
The spacetime has time-translational and axial Killing vectors $t^\mu \partial_\mu=\partial_t$ and $\varphi^\mu \partial_\mu=\partial_\varphi$ 
because of its stationarity and axisymmetry, respectively.
We can assume $\vartheta=\pi/2$ without loss of generality because of spherical symmetry of the spacetime.

\begin{figure}[htbp]
\begin{center}
\includegraphics[width=85mm]{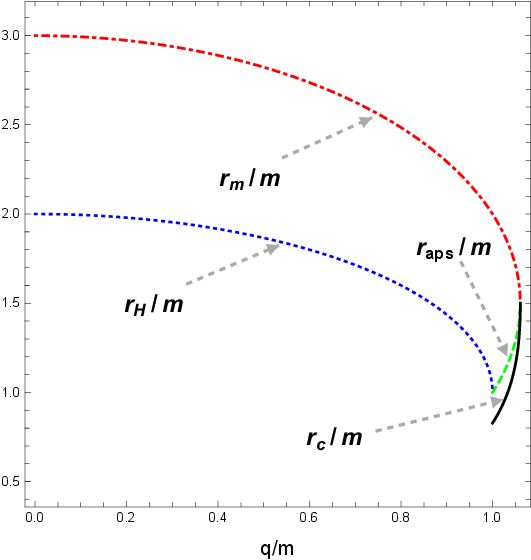}
\end{center}
\caption{Specific radial coordinates of $r_\mathrm{m}/m$, $r_\mathrm{aps}/m$, $r_\mathrm{H}/m$, and $r_\mathrm{c}/m$.
A (red) dot-dashed,
a (green) dashed, 
a (blue) dotted,  
and a (black) solid curves 
denote the photon sphere at $3/2 \leq r_\mathrm{m}/m \leq 3$ for $0 \leq q/m \leq 3/(2\sqrt{2})$,
the antiphoton sphere at $1 < r_{\mathrm{aps}}/m \leq 3/2$ for $1 < q/m \leq  3/(2\sqrt{2})$,
the event horizon at $1 \leq r_\mathrm{H}/m \leq 2$ for $0 \leq q/m \leq 1$,
and the smaller positive zero point of an effective potential at $2(\sqrt{2}-1)<r_\mathrm{c}/m \leq 3/2$ for $1< q/m \leq  3/(2\sqrt{2})$, respectively.
}
\label{fig:1}
\end{figure}

From $k^\mu k_\mu=0$, where $k^\mu\equiv \dot{x}^\mu$ is a wave vector and the dot denotes a differentiation with respect to an affine parameter, 
the trajectory of a light is given by
\begin{equation}\label{eq:3}
-A\dot{t}^2+\frac{\dot{r}^2}{A}+r^2 \dot{\varphi}^2=0.
\end{equation}
The light at the closest distant $r=r_0$ satisfies
\begin{equation}\label{eq:4}
A_0\dot{t}^2_0=r^2_0 \dot{\varphi}^2_0,
\end{equation}
Here and hereinafter, quantities with the subscript $0$ denote the quantities at $r=r_0$.
The impact parameter on the light is given by
\begin{equation}
b(r_0)\equiv \frac{L}{E}=\frac{r_0^2 \dot{\varphi}_0}{A_0\dot{t}_0},
\end{equation}
where $E\equiv -g_{\mu \nu} t^\mu k^\nu=A\dot{t}$ and 
$L \equiv g_{\mu \nu} \varphi^\mu k^\nu=r^2 \dot{\varphi}$ 
are the conserved energy and angular momentum of the light ray, respectively.
By using Eq.~(\ref{eq:4}),
it can be rewritten as
\begin{equation}
b=\pm \sqrt{\frac{r_0^2}{A_0}}.
\end{equation}
Note that $b$, $E$, and $L$ are constant along the trajectory of the ray.
We concentrate on the positive impact parameter unless we say the negative impact parameter.
Equation~(\ref{eq:3}) is expressed by
$\dot{r}^2+V(r)/E^2=0$,
where $V(r)$ is the effective potential of the light defined by
\begin{equation}
V(r)\equiv \frac{Ab^2}{r^2}-1.
\end{equation}
The light can be in the nonpositive region of the effective potential $V(r)\leq 0$.
The larger and smaller positive solutions of the equation $V^\prime=0$, where the prime is a differentiation with respect to the radial coordinate $r$,
are $r=r_\mathrm{m}$ for $q\leq 3m/(2\sqrt{2})$ and $r=r_\mathrm{aps}$ for $m<q\leq 3m/(2\sqrt{2})$, 
where $r_\mathrm{m}$ and $r_\mathrm{aps}$ are given by
\begin{equation}\label{eq:5}
r_\mathrm{m}\equiv\frac{3m+\sqrt{9m^{2}-8q^{2}}}{2}
\end{equation}
and 
\begin{equation}
r_\mathrm{aps}\equiv \frac{3m-\sqrt{9m^{2}-8q^{2}}}{2},
\end{equation}
respectively.
Note that $r=r_\mathrm{m}$ and $r=r_\mathrm{aps}$ hold an equation
\begin{equation}
r^{2}-3mr+2q^{2}=0.
\end{equation}
We name an impact parameter $b\left(r_0=r_\mathrm{m}\right)=b_\mathrm{m}$ which satisfies $V_\mathrm{m}=0$ critical impact parameter. 
Here and hereinafter, quantities with the subscript m denote the quantities at $r=r_\mathrm{m}$ or $r_0=r_\mathrm{m}$.
A circular light orbit with the critical impact parameter $b_\mathrm{m}$ at $r=r_\mathrm{m}$ for $q< 3m/(2\sqrt{2})$ is unstable
because of $V_\mathrm{m}=V^{\prime}_\mathrm{m}=0$ and $V^{\prime \prime}_\mathrm{m}<0$ 
and the sphere of the unstable circular light orbit is called photon sphere. 
On the other hand, a circular light orbit with an impact parameter which holds 
$V(r_\mathrm{aps})=V^{\prime}(r_\mathrm{aps})=0$ for $m<q< 3m/(2\sqrt{2})$ is stable since  
$V^{\prime \prime}(r_\mathrm{aps})>0$ holds. 
The sphere of the stable circular light orbit is called antiphoton sphere. 
For a marginal case $q=3m/(2\sqrt{2})$, 
the light ray with the critical impact parameter $b_\mathrm{m}$ holds 
$V_\mathrm{m}=V^{\prime}_\mathrm{m}=V^{\prime \prime}_\mathrm{m}=0$ and $V^{\prime \prime \prime}_\mathrm{m}<0$ and
the photon sphere and the antiphoton sphere degenerate to form a marginally unstable photon sphere at $r=r_\mathrm{m}=r_\mathrm{aps}=3m/2$. 
The radii of the photon sphere $r_\mathrm{m}$ and the antiphoton sphere $r_\mathrm{aps}$ are shown in Fig.~\ref{fig:1}.

From Eq.~(\ref{eq:3}), we obtain the deflection angle $\alpha$ of the light as 
\begin{equation}\label{eq:defal}
\alpha=I(r_0)-\pi,
\end{equation}
where $I(r_0)$ is given by
\begin{equation}\label{eq:8}
I(r_0)\equiv 2 \int^\infty_{r_0} \frac{b \mathrm{d}r}{r^2 \sqrt{-V(r)}}.
\end{equation}

\section{Deflection angle in strong deflection limits}
In this section, we show that parameters $\bar{a}$ and $\bar{b}$ in the deflection angle~(\ref{eq:1}) for $q/m<3/(2\sqrt{2})$
and parameters $\bar{c}$ and $\bar{d}$ in the deflection angle~(\ref{eq:2}) for $1<q/m<3/(2\sqrt{2})$ in the strong deflection limits
in the Reissner-Nordstr\"{o}m black hole and naked singularity spacetimes
can be obtained analytically without Taylor expansions on the electrical charge.

\subsection{Light rays barely outside of the photon sphere}
\begin{figure}[htbp]
\begin{center}
\includegraphics[width=85mm]{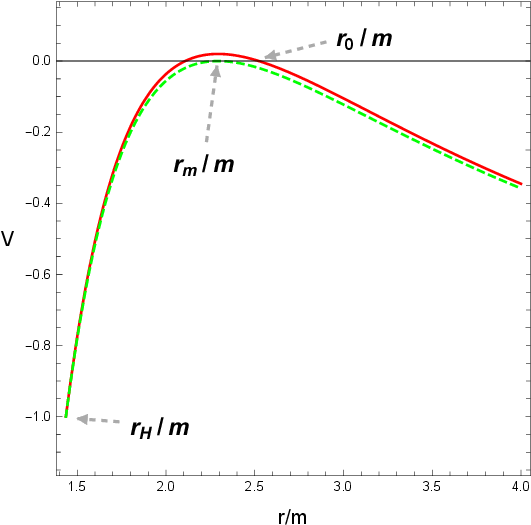}
\end{center}
\caption{Effective potential~$V$ of a ray with $b=1.01 b_\mathrm{m}=4.36m$ to pass slightly outside of the photon sphere 
around a black hole of~$q=0.9 m$ with an event horizon at $r_\mathrm{H}/m=1.44$ is shown as a solid (red) curve. 
The closest distance of the light is $r_\mathrm{0}/m=2.52$. 
A dashed (green) curve denotes the effective potential~$V$ with the photon sphere at $r_\mathrm{m}/m=2.29$ in the critical case $b=b_\mathrm{m}=4.32m$.}
\label{fig:2}
\end{figure}

\begin{figure}[htbp]
\begin{center}
\includegraphics[width=85mm]{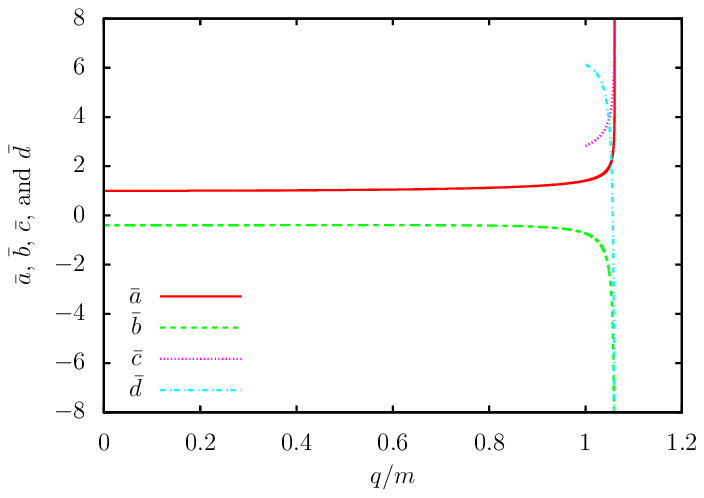}
\end{center}
\caption{Parameters $\bar{a}$ and $\bar{b}$ in the deflection angle~(\ref{eq:1}) 
and parameters $\bar{c}$ and $\bar{d}$ in the deflection angle~(\ref{eq:2}) in the strong deflection limits.  
Solid~(red), dashed~(green), dotted~(magenta), and dot-dashed~(cyan) curves denote $\bar{a}$, $\bar{b}$, $\bar{c}$, and $\bar{d}$, respectively.
}
\label{fig:3}
\end{figure}

\begin{figure}[htbp]
\begin{center}
\includegraphics[width=85mm]{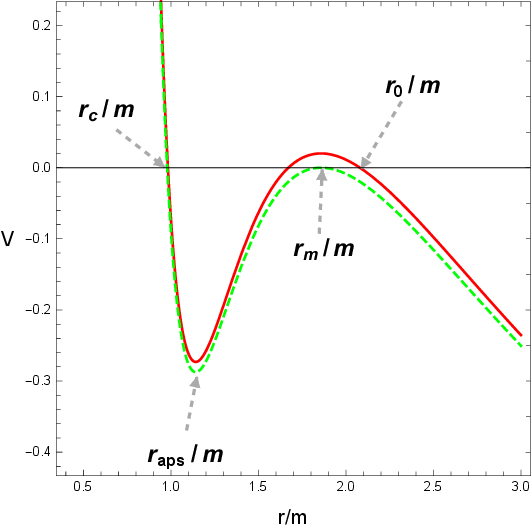}
\end{center}
\caption{Effective potential~$V$ of a ray with $b/m=1.01 b_\mathrm{m}/m=3.91$ to pass little outside of the photon sphere at $r_\mathrm{m}/m=1.86$
around the naked singularity with~$q/m=1.03$ is shown as a solid (red) curve. 
Its reflectional point is at $r_\mathrm{0}/m=2.08$.
A dashed (green) curve denotes the effective potential~$V$ of the critical case with $b/m=b_\mathrm{m}/m=3.87$ 
and its smaller positive root is at $r_\mathrm{c}/m=0.979$ and an antiphoton sphere is at $r_\mathrm{aps}/m=1.14$.}
\label{fig:4}
\end{figure}

We consider light rays to form images little outside of the photon sphere in the Reissner-Nordstr\"{o}m spacetime for $q/m<3/(2\sqrt{2})$. 
Its effective potential is shown in Fig.~\ref{fig:2}.
The analytic expressions of $\bar{a}$ and $\bar{b}$ of the deflection angle~(\ref{eq:1})
in a strong deflection limit $r_0 \rightarrow r_\mathrm{m}+0$ or $b \rightarrow b_\mathrm{m}+0$ are obtained as~\cite{Tsukamoto:2016jzh}, 
\begin{equation}
\bar{a}=\frac{r_\mathrm{m}}{\sqrt{3mr_\mathrm{m}-4q^{2}}}
\end{equation}
and
\begin{eqnarray}
\bar{b}
&=&\bar{a} \log \left[ \frac{8(3mr_\mathrm{m}-4q^{2})^{3}}{m^{2}r_\mathrm{m}^{2}(mr_\mathrm{m}-q^{2})^{2}} \right. \nonumber\\
&&\left. \times \left(2\sqrt{mr_\mathrm{m}-q^{2}}-\sqrt{3mr_\mathrm{m}-4q^{2}} \right)^{2} \right] -\pi, \nonumber\\
\end{eqnarray}
respectively.
The parameters recover numerical calculations by Eiroa \textit{et al.}~\cite{Eiroa:2002mk} and partly numerical calculations by Bozza~\cite{Bozza:2002zj} 
as shown in Ref.~\cite{Tsukamoto:2016jzh}
and they are shown in Fig.~\ref{fig:3}.
Notice that the analytic expressions of $\bar{a}$ and $\bar{b}$ are valid also in the Reissner-Nordstr\"{o}m naked singularity spacetime with $1<q/m<3/(2\sqrt{2})$.
The effective potential of the light to pass barely outside of the photon sphere is shown in Fig.~\ref{fig:4}.

\subsection{Light rays little inside of the photon sphere}
\begin{figure}[htbp]
\begin{center}
\includegraphics[width=85mm]{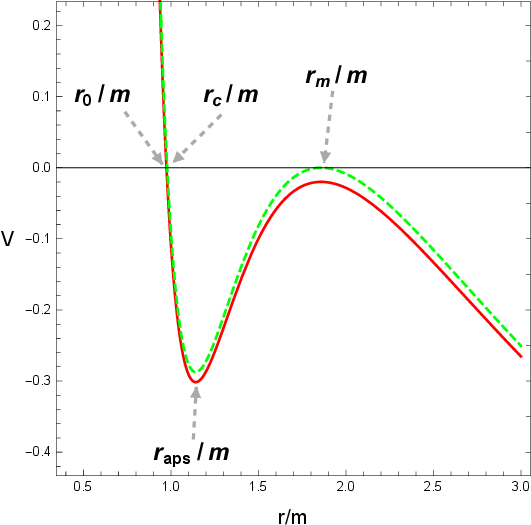}
\end{center}
\caption{Effective potential~$V$ of a ray with $b=0.99 b_\mathrm{m}=3.82m$ to pass slightly inside of the photon sphere
around the naked singularity is shown as a solid (red) curve. 
Its reflectional point is at $r_\mathrm{0}/m=0.975$.
The dashed (green) curve and the values of $q/m$, $b_\mathrm{m}/m$, $r_\mathrm{c}/m$, $r_\mathrm{m}/m$, and $r_\mathrm{aps}/m$
are the same as Fig.~\ref{fig:4}.}
\label{fig:5}
\end{figure}

In the Reissner-Nordstr\"{o}m naked singularity spacetime with $1<q/m<3/(2\sqrt{2})$,
images not only barely outside but also inside of the photon sphere can appear 
and its effective potential is shown in Fig.~\ref{fig:5}. 
As following Refs.~\cite{Shaikh:2019itn,Tsukamoto:2021caq}, we consider the deflection angle of a light ray 
which is reflected a potential barrier near an antiphoton sphere in a strong deflection limit $r_0 \rightarrow r_\mathrm{c}$, 
where $r=r_\mathrm{c}$ is the smaller positive root of the effective potential with the critical impact parameter $b=b_\mathrm{m}$.
We note $b_\mathrm{c}\equiv b(r_\mathrm{c})=b_\mathrm{m}$.
Here and hereinafter, quantities with the subscript $c$ denote the quantities at $r_0=r_\mathrm{c}$.
Note that $r_\mathrm{c}$ satisfies an equation  
\begin{equation}
\frac{A_\mathrm{m}}{r_\mathrm{m}^2}r_\mathrm{c}^4-r_\mathrm{c}^2+2mr_\mathrm{c}-q^2=0
\end{equation}
and $r_\mathrm{c}$ can be obtained analytically as 
\begin{equation}\label{eq:9}
r_\mathrm{c}=\frac{r_\mathrm{m}\left( \sqrt{mr_\mathrm{m}}-\sqrt{mr_\mathrm{m}-q^2} \right)}{\sqrt{mr_\mathrm{m}-q^2}}.
\end{equation}
Here, $A_\mathrm{m}$ is given by
\begin{equation}
A_\mathrm{m}=\frac{1}{3} \left( 1-\frac{q^2}{r_\mathrm{m}^2} \right)=\frac{m}{r_\mathrm{m}}-\frac{q^2}{r_\mathrm{m}^2}>0.
\end{equation}

By using a variable 
\begin{equation}
z\equiv 1-\frac{r_\mathrm{m}}{r},
\end{equation}
Eq.~(\ref{eq:8}) can be written in
\begin{equation}
I(r_0)=\int^1_{\gamma(r_0)} f(z,r_0)\mathrm{d}z,
\end{equation}
where 
\begin{equation}
\gamma(r_0)\equiv 1-\frac{r_\mathrm{m}}{r_0}
\end{equation}
and
\begin{eqnarray}
f(z,r_0) \equiv \frac{2}{\sqrt{h(z,r_0)}},
\end{eqnarray}
and where $h(z,r_0)$ is defined by
\begin{equation}
h(z,r_0)=c_1(r_0)+c_2z^2+c_3z^3+c_4z^4,
\end{equation}
where $c_1(r_0)$, $c_2$, $c_3$, and $c_4$ are given by
\begin{eqnarray}
c_1(r_0)&\equiv& A_\mathrm{m} \left( \frac{b_\mathrm{m}^2}{b^2}-1 \right),\\
c_2&\equiv& 1-\frac{2q^2}{r^2_\mathrm{m}}=\frac{3m}{r_\mathrm{m}}-\frac{4q^2}{r_\mathrm{m}^2},\\
c_3&\equiv& -\frac{2}{3} \left( 1-\frac{4q^2}{r^2_\mathrm{m}} \right)=-\frac{2m}{r_\mathrm{m}}+\frac{4q^2}{r_\mathrm{m}^2}, \\
c_4&\equiv& -\frac{q^2}{r^2_\mathrm{m}}.
\end{eqnarray}

In the strong deflection limit $r_0 \rightarrow r_\mathrm{c}-0$ or $b \rightarrow b_\mathrm{c}-0=b_\mathrm{m}-0$, we obtain 
\begin{eqnarray}
c_1(r_0)&\rightarrow& +0, \\
\gamma(r_0)&\rightarrow&1-\frac{r_\mathrm{m}}{r_\mathrm{c}}<0.
\end{eqnarray}
We expand $b(r_0)$ in powers of $r_0-r_{\mathrm{c}}<0$ as 
\begin{equation}
b(r_0)=b_{\mathrm{c}}+b^{\prime}_{\mathrm{c}} \left( r_0-r_{\mathrm{c}} \right)+O\left( \left( r_0-r_{\mathrm{c}} \right)^2 \right),
\end{equation}
where $b_{\mathrm{c}}$ and $b^{\prime}_{\mathrm{c}}$ are given by 
\begin{eqnarray}
b_{\mathrm{c}}=b_\mathrm{m}=\frac{r_\mathrm{m}}{\sqrt{A_\mathrm{m}}}=\frac{r_\mathrm{m}^2}{\sqrt{mr_\mathrm{m}-q^2}},
\end{eqnarray}
and 
\begin{equation}
b_\mathrm{c}^\prime=A_\mathrm{c}^{-\frac{3}{2}} \left(1-\frac{3m}{r_\mathrm{c}} +\frac{2q^2}{r_\mathrm{c}^2} \right),
\end{equation}
respectively.
We separate $I(r_0)$ into a divergent part
$I_\mathrm{D}$ and a regular part $I_\mathrm{R}$.
The divergent part $I_\mathrm{D}$ is defined by
\begin{equation}
I_\mathrm{D}\equiv \int^1_{\gamma(r_0)} f_\mathrm{D}(z,r_0)\mathrm{d}z,
\end{equation}
where $f_\mathrm{D}(z,r_0)$ is defined by
\begin{equation}
f_\mathrm{D}(z,r_0) \equiv \frac{2}{\sqrt{c_1(r_0)+c_2z^2}}.
\end{equation}
It can be integrated as 
\begin{equation}
I_\mathrm{D}= \frac{2}{\sqrt{c_2}} \log \frac{c_2+\sqrt{c_2(c_1+c_2)}}{c_2\gamma+\sqrt{c_2\left( c_1+c_2\gamma^2\right)}}.
\end{equation}
Here, we have used $c_2>0$ for $1\leq q/m <3/(2\sqrt{2})$.

By using approximations 
\begin{equation}
\sqrt{c_{2}\left( c_{1\mathrm{c}}+c_{2}\gamma^2_{\mathrm{c}}\right)}
\sim -c_{2} \gamma_{\mathrm{c}} \left( 1+ \frac{c_{1\mathrm{c}}}{2 c_{2} \gamma^2_{\mathrm{c}}} \right),
\end{equation}
we obtain the divergent part $I_\mathrm{D}$ in the strong deflection limit~$r_0 \rightarrow r_\mathrm{c}-0$ or $b \rightarrow b_\mathrm{c}-0=b_{\mathrm{m}}-0$ as
\begin{eqnarray}
I_\mathrm{D}
&=&\bar{c} \log \left( -\frac{4c_{2} \gamma_\mathrm{c}}{c_{1\mathrm{c}}} \right) \nonumber\\
&=&-\bar{c} \log \left( \frac{b_\mathrm{m}^2}{b^2}-1 \right) \nonumber\\
&&+\bar{c} \log \left[ \frac{4\left( 3mr_\mathrm{m}-4q^2 \right)}{mr_\mathrm{m}-q^2} \left( \frac{r_\mathrm{m}}{r_\mathrm{c}}-1 \right)  \right],
\end{eqnarray}
where $\bar{c}$ is given by
\begin{equation}
\bar{c}\equiv \frac{2r_\mathrm{m}}{\sqrt{3mr_\mathrm{m}-4q^2}}.
\end{equation}

The regular part $I_\mathrm{R}$ is given by
\begin{equation}
I_\mathrm{R}(r_0)\equiv \int^1_{\gamma(r_0)} g(z,r_0) \mathrm{d}z, 
\end{equation}
where $g(z,r_0)$ is defined by
\begin{equation}
g(z,r_0)\equiv f(z,r_0)-f_\mathrm{D}(z,r_0).
\end{equation}
The regular part $I_\mathrm{R}$ can be expanded as, in powers of $r_0-r_{\mathrm{c}}$, 
\begin{equation}
I_\mathrm{R}(r_0)=\sum^\infty_{j=0} \frac{1}{j!}(r_0-r_{\mathrm{c}})^j \int^1_{\gamma_\mathrm{c}} \left. \frac{\partial^j g}{\partial r_0^j} \right|_{r_0=r_{\mathrm{c}}} \mathrm{d}z.
\end{equation}
We are interested in the term with $j=0$ and we obtain $I_\mathrm{R}$ as
\begin{eqnarray}
I_\mathrm{R}
&=&\int^1_{\gamma_\mathrm{c}} g(z,r_{\mathrm{c}}) \mathrm{d}z \nonumber\\
&=&\int^1_{\gamma_\mathrm{c}} \left( \frac{2}{ \sqrt{c_{2}+c_{3}z+c_{4}z^2} \left| z \right| }- \frac{2}{  \sqrt{c_{2}} \left| z \right| }  \right) \mathrm{d}z. \nonumber\\
\end{eqnarray}
We note ${\gamma_\mathrm{c}}<0$ and we can integrate $I_\mathrm{R}$ as 
\begin{eqnarray}
I_\mathrm{R}
&=&\int^0_{\gamma_\mathrm{c}} \left( \frac{-2}{ \sqrt{c_{2}+c_{3}z+c_{4}z^2} z }+ \frac{2}{  \sqrt{c_{2}}z}  \right) \mathrm{d}z \nonumber\\
&&+\int^1_{0} \left( \frac{2}{ \sqrt{c_{2}+c_{3}z+c_{4}z^2} z }- \frac{2}{ \sqrt{c_{2}} z }  \right) \mathrm{d}z \nonumber\\
&=&\bar{c} \log \left[ \frac{16c_2^2}{c_3 \gamma_\mathrm{c} +2c_2 +2\sqrt{c_2 \left( c_2+c_3 \gamma_\mathrm{c} +c_4 \gamma_\mathrm{c}^2 \right)} } \right. \nonumber\\
&&\qquad \qquad \times \left. \frac{1}{c_3 +2c_2 +2\sqrt{c_2 \left( c_2 +c_3 +c_4 \right)} } \right] \nonumber\\
&=&\bar{c} \log \left[ \frac{4\left(3 m r_\mathrm{m}-4 q^2\right)^2 r_\mathrm{c}} {2(mr_\mathrm{m}-q^2)+\sqrt{(3mr_\mathrm{m}- 4q^2)(mr_\mathrm{m}-q^2)}         } \right. \nonumber\\
&&\qquad \qquad \times \left. \frac{1}{mr_\mathrm{m}(r_\mathrm{m}+2r_\mathrm{c})-2q^2(r_\mathrm{m}+r_\mathrm{c})+\sqrt{G}} \right], \nonumber\\
\end{eqnarray}
where 
\begin{eqnarray}
G=\left(3 m r_\mathrm{m}-4 q^2\right) \left[m r_\mathrm{m} r_\mathrm{c} (2 r_\mathrm{m}+r_\mathrm{c})-q^2 (r_\mathrm{m}+r_\mathrm{c})^2 \right].\nonumber\\
\end{eqnarray}
From Eqs.~(\ref{eq:5}) and~(\ref{eq:9}), we get $G=0$ and we obtain $I_\mathrm{R}$ as
\begin{eqnarray}
I_\mathrm{R}
&=&\bar{c} \log \left[ \frac{4\left(3 m r_\mathrm{m}-4 q^2\right)^2 r_\mathrm{c}} {2(mr_\mathrm{m}-q^2)+\sqrt{(3mr_\mathrm{m}- 4q^2)(mr_\mathrm{m}-q^2)}         } \right. \nonumber\\
&&\qquad \qquad \times \left. \frac{1}{mr_\mathrm{m}(r_\mathrm{m}+2r_\mathrm{c})-2q^2(r_\mathrm{m}+r_\mathrm{c})} \right]. \nonumber\\
\end{eqnarray}
Therefore, the term $\bar{d}$ in the strong deflection limit $r_0 \rightarrow r_\mathrm{c}-0$ or $b \rightarrow b_\mathrm{m}-0$ is given by, in the following analytic form, 
\begin{eqnarray}
\bar{d}
&=&\bar{c} \log \left[ \frac{16\left(3 m r_\mathrm{m}-4 q^2\right)^3 (r_\mathrm{m}-r_\mathrm{c})}{2(mr_\mathrm{m}-q^2)+\sqrt{(3mr_\mathrm{m}- 4q^2)(mr_\mathrm{m}-q^2)}         } \right. \nonumber\\
&&\times \left. \frac{1}{(mr_\mathrm{m}-q^2) \left\{ mr_\mathrm{m}(r_\mathrm{m}+2r_\mathrm{c})-2q^2(r_\mathrm{m}+r_\mathrm{c}) \right\} } \right] -\pi. \nonumber\\
\end{eqnarray}
Figure~6 shows the percent error of deflection angle calculated by 
\begin{eqnarray}\label{eq:per}
\frac{\alpha \mathrm{\:  of \: Eq. \: (1.2)}-\alpha \mathrm{\: of \: Eq. \: (2.12)}}{\alpha \mathrm{ \:of \: Eq. \: (2.12)}} \times 100 
\end{eqnarray}
as a function of $\alpha$ of Eq.~(\ref{eq:defal}). It has confirmed the percent error in the case of $q^2/m^2=1.05$ shown as Fig.~3 in Ref.~\cite{Shaikh:2019itn}.
\begin{figure}[htbp]
\begin{center}
\includegraphics[width=85mm]{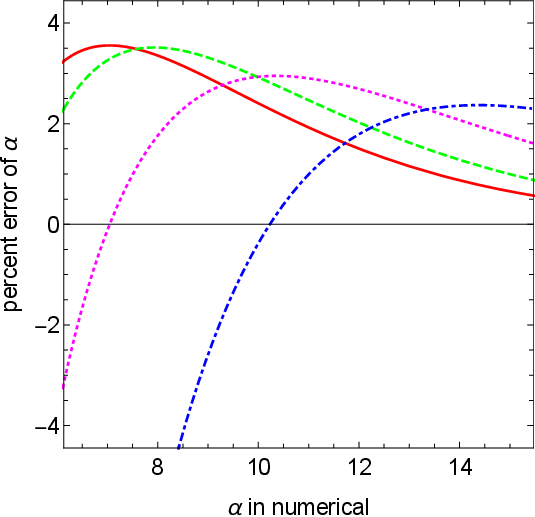}
\end{center}
\caption{The percent error of deflection angle defined by Eq.~(\ref{eq:per}) as a function of $\alpha$ of Eq.~(\ref{eq:defal}).
A solid (red), dashed (green), dotted (magenta), and dot-dashed (blue) curves denote the percent error 
for $q/m=1.01$, $\sqrt{1.05}\sim 1.025$, $1.04$, and $1.05$, respectively.}
\label{fig:6}
\end{figure}

\section{Observables in strong deflection limits}
\begin{figure}[htbp]
\begin{center}
\includegraphics[width=85mm]{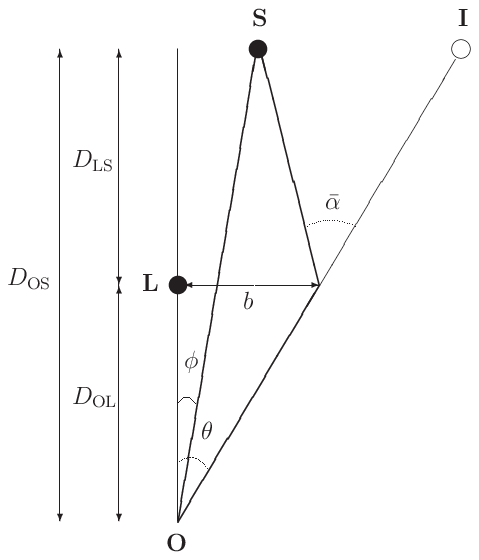}
\end{center}
\caption{Usual configuration of gravitational lensing. 
A light ray having an impact parameter~$b$ emitted by a source S at a source angle~$\phi$ 
is reflected with an effective deflection angle $\bar{\alpha}$ by a lens L 
and the ray is observed as an image I at an image angle $\theta$ by an observer O. 
$D_{\mathrm{OS}}$, $D_{\mathrm{LS}}$, and $D_{\mathrm{OL}}$ denote 
the distances between O and S, between L and S, and between O and L, respectively.
We assume that all the angles $\phi$, $\theta$, and $\bar{\alpha}$ are small.}
\label{fig:7}
\end{figure}

We consider a usual gravitational lens configuration, as shown Fig.~\ref{fig:7},
that a source S at a source angle $\phi$ emits a ray having an impact parameter $b$,
it is deflected by a lens L at an deflection angle $\alpha$, 
and that an observer O sees its image I with an image angle $\theta$.
We assume that the angles are small, i.e., $\bar{\alpha}\ll 1$, $\theta=b/D_{\mathrm{OL}}\ll 1$, and $\phi \ll 1$,
where $D_{\mathrm{OL}}$ is a distance between O and L and $\bar{\alpha}$ is an effective deflection angle given by
\begin{equation}
\bar{\alpha}=\alpha \quad  \mathrm{mod} \quad  2\pi.
\end{equation}
By using the winding number $N$ of the light, the deflection angle~$\alpha$ is obtained as
\begin{equation}\label{eq:10}
\alpha=\bar{\alpha}+ 2\pi N.
\end{equation}
We use a small lens equation~\cite{Bozza:2008ev} given by
\begin{equation}\label{eq:11}
D_{\mathrm{LS}}\bar{\alpha}=D_{\mathrm{OS}} \left( \theta-\phi \right),
\end{equation}
where $D_{\mathrm{LS}}$ and $D_{\mathrm{OS}}=D_{\mathrm{OL}}+D_{\mathrm{LS}}$ 
are distances between L and S and between O and S, respectively.
We expand the deflection angle $\alpha(\theta)$ around $\theta=\theta^{0}_{N}$ as  
\begin{equation}\label{eq:12}
\alpha(\theta)=\alpha(\theta^{0}_{N})+\left. \frac{\mathrm{d}\alpha}{\mathrm{d}\theta} \right|_{\theta=\theta^{0}_{N}} (\theta-\theta^{0}_{N})+O\left( (\theta-\theta^{0}_{N})^2 \right),
\end{equation}
where $\theta^{0}_{N}$ defined by
\begin{equation}\label{eq:13}
\alpha(\theta^{0}_{N}) = 2\pi N.
\end{equation}

\subsection{Images barely outside of the photon sphere}
By following Refs.~\cite{Bozza:2002zj,Tsukamoto:2021caq},
we calculate images little outside of the photon sphere. 
The deflection angle $\alpha$ in a strong deflection limit~$b\rightarrow b_{\mathrm{m}}+0$ is given by  
\begin{eqnarray}\label{eq:14}
\alpha(\theta)
&=&-\bar{a} \log \left( \frac{\theta}{\theta_\infty}-1 \right)+\bar{b} \nonumber\\
&&+O\left( \left( \frac{\theta}{\theta_\infty}-1 \right) \log \left( \frac{\theta}{\theta_\infty}-1 \right) \right),
\end{eqnarray}
where $\theta_\infty \equiv b_{\mathrm{m}}/D_{\mathrm{OL}}$ is the image angle of the photon sphere.
From Eqs.~(\ref{eq:13}) and (\ref{eq:14}), we rewrite $\theta^{0}_{N}$ as 
\begin{equation}\label{eq:15}
\theta^{0}_{N}=\left( 1+\exp \left( {\frac{\bar{b}-2\pi N}{\bar{a}}} \right) \right) \theta_\infty.
\end{equation}
By using $\left. d\alpha /d\theta \right|_{\theta=\theta^{0}_{N}}=\bar{a}/(\theta_\infty-\theta^{0}_{N})$
and Eqs.~(\ref{eq:10}), (\ref{eq:12}), (\ref{eq:13}), and (\ref{eq:15}),
the effective deflection angle $\bar{\alpha}(\theta_{N})$ for the positive solution $\theta=\theta_{N}$ of the lens equation for a positive winding number~$N$
is given by
\begin{equation}\label{eq:16}
\bar{\alpha}(\theta_{N})=\frac{\bar{a}\left( \theta^{0}_{N}-\theta_{N} \right)}{\theta_\infty \exp \left(\frac{\bar{b}-2\pi N}{\bar{a}} \right)}.
\end{equation}
From Eqs.~(\ref{eq:11}) and (\ref{eq:16}), we get the image angle as
\begin{equation}
\theta_{N}(\phi) \sim \theta^{0}_{N} - \frac{\theta_\infty D_{\mathrm{OS}} (\theta^{0}_{N}-\phi)\exp\left(\frac{\bar{b}-2\pi N}{\bar{a}}\right) }{\bar{a}D_{\mathrm{LS}}}
\end{equation}
and its magnification $\mu_{N}$ as 
\begin{eqnarray}
\mu_{N} 
&\equiv& \frac{\theta_{N}}{\phi}\frac{\mathrm{d}\theta_{N}}{\mathrm{d}\phi} \nonumber\\
&\sim& \frac{\theta^2_\infty D_{\mathrm{OS}} \left(1+ \exp\left(\frac{\bar{b}-2\pi N}{\bar{a}}\right)\right) \exp\left({\frac{\bar{b}-2\pi N}{\bar{a}}}\right)}{\phi \bar{a} D_{\mathrm{LS}}}. \nonumber\\
\end{eqnarray}
The image angle $\theta_{\mathrm{E}N}$ of the relativistic Einstein ring is obtained as
\begin{equation}
\theta_{\mathrm{E}N}\equiv \theta_{N}(0) \sim \left( 1- \frac{\theta_\infty D_{\mathrm{OS}} \exp\left(\frac{\bar{b}-2\pi N}{\bar{a}}\right)}{\bar{a}D_{\mathrm{LS}}} \right) \theta^{0}_{N}.
\end{equation}
The difference of the image angles of the photon sphere and the outermost image is given by 
\begin{equation}
\bar{\mathrm{s}}\equiv \theta_1-\theta_\infty \sim \theta^{0}_1-\theta^{0}_\infty = \theta_\infty \exp\left(\frac{\bar{b}-2\pi}{\bar{a}}\right). 
\end{equation}
We obtain the sum of the magnifications of all the images as
\begin{eqnarray}
\sum^\infty_{N=1} \mu_{N} \sim \frac{\theta^2_\infty D_{\mathrm{OS}} \left(1+\exp\left(\frac{2\pi}{\bar{a}}\right) +\exp\left(\frac{\bar{b}}{\bar{a}}\right)\right) \exp\left(\frac{\bar{b}}{\bar{a}}\right)}{\phi \bar{a} D_{\mathrm{LS}}\left( \exp\left(\frac{4\pi}{\bar{a}}\right) -1 \right)} \nonumber\\
\end{eqnarray}
and the one of images excluding the outermost image as
\begin{eqnarray}
&&\sum^\infty_{N=2} \mu_{N} \sim \nonumber\\
&&\frac{\theta^2_\infty D_{\mathrm{OS}} \left(\exp\left(\frac{2\pi}{\bar{a}}\right)+\exp\left(\frac{4\pi}{\bar{a}}\right) +\exp\left(\frac{\bar{b}}{\bar{a}}\right)\right) \exp\left(\frac{\bar{b}-4\pi}{\bar{a}}\right)}{\phi \bar{a} D_{\mathrm{LS}}\left( \exp\left(\frac{4\pi}{\bar{a}}\right) -1 \right)}.\nonumber\\
\end{eqnarray}
The ratio of the magnifications of the outermost image to the sum of the other images is given by
\begin{equation}
\bar{\mathrm{r}} 
\equiv \frac{\mu_1}{\sum^\infty_{N=2} \mu_{N}}
\sim \frac{\left( \exp\left(\frac{4\pi}{\bar{a}}\right)-1 \right) \left( \exp\left(\frac{2\pi}{\bar{a}}\right)+\exp\left(\frac{\bar{b}}{\bar{a}}\right) \right)}{ \exp\left(\frac{2\pi}{\bar{a}}\right)+\exp\left(\frac{4\pi}{\bar{a}}\right) +\exp\left(\frac{\bar{b}}{\bar{a}}\right) }.
\end{equation}

\subsection{Images little inside of the photon sphere}
We calculate observables in the strong deflection limit~$b\rightarrow b_{\mathrm{m}}-0$ as well as Refs.~\cite{Shaikh:2019itn,Tsukamoto:2021caq}.
The deflection angle $\alpha$ is expressed by
\begin{eqnarray}\label{eq:17}
\alpha(\theta)
&=&-\bar{c} \log \left( \frac{\theta_\infty^2}{\theta^2}-1 \right)+\bar{d} \nonumber\\
&&+O\left( \left( \frac{\theta_\infty}{\theta}-1 \right) \log \left( \frac{\theta_\infty}{\theta}-1 \right) \right).
\end{eqnarray}
We obtain $\theta^{0}_{N}$, from Eqs.~(\ref{eq:13}) and (\ref{eq:17}),
\begin{equation}\label{eq:18}
\theta^{0}_{N}=\frac{\theta_\infty}{\sqrt{1+e_N}},
\end{equation}
where $e_N$ is defined by
\begin{equation}
e_N\equiv \exp\left( \frac{\bar{d}-2\pi N}{\bar{c}}\right).
\end{equation}
From 
\begin{equation}
\left. \frac{\mathrm{d}\alpha}{\mathrm{d}\theta} \right|_{\theta=\theta^{0}_{N}}= \frac{2\bar{c}\theta_\infty^2}{\theta^{0}_{N}(\theta_\infty-\theta^{0}_{N})(\theta_\infty+\theta^{0}_{N})}
\end{equation}
and Eqs.~(\ref{eq:10}), (\ref{eq:12}), (\ref{eq:13}), and (\ref{eq:18}), 
the effective deflection angle $\bar{\alpha}(\theta_{N})$ for the positive solution $\theta=\theta_{N}$ of the lens equation for a positive winding number $N$ 
is obtained as 
\begin{equation}\label{eq:19}
\bar{\alpha}(\theta_{N})=\frac{2\bar{c}(1+e_N)^\frac{3}{2} \left( \theta_{N}-\theta^{0}_{N} \right)}{\theta_\infty e_N}.
\end{equation}
We obtain the solution $\theta_{N}(\phi)$ as, from Eqs.~(\ref{eq:11}) and (\ref{eq:19}), 
\begin{equation}
\theta_{N}(\phi) \sim \theta^{0}_{N} + \frac{\theta_\infty D_{\mathrm{OS}} e_N (\theta^{0}_{N}-\phi)}{2\bar{c}D_{\mathrm{LS}} (1+e_N)^\frac{3}{2} }
\end{equation}
and its magnification is given by
\begin{equation}
\mu_{N} \sim -\frac{\theta^2_\infty D_{\mathrm{OS}} e_N }{2 \phi \bar{c} D_{\mathrm{LS}}(1+e_N)^2}.
\end{equation}
The relativistic Einstein ring angle is given by
\begin{equation}
\theta_{\mathrm{E}N}\sim \left( 1+ \frac{\theta_\infty D_{\mathrm{OS}}e_N}{2\bar{c}D_{\mathrm{LS}}(1+e_N)^\frac{3}{2}} \right) \theta^{0}_{N}.
\end{equation}
The difference of the image angles between the photon sphere and the innermost image is given by 
\begin{equation}
\bar{\mathrm{s}}= \theta_1-\theta_\infty \sim \theta^{0}_1-\theta^{0}_\infty = \frac{1-\sqrt{1+e_1}}{\sqrt{1+e_1}} \theta_\infty.
\end{equation}

\section{Conclusion}
On this paper, we investigate gravitational lensing by a photon sphere in a Reissner-Nordstr\"{o}m naked singularity spacetime.
Infinite numbers of images little not only inside but also outside of the photon sphere can be formed
because of a potential barrier near an antiphoton sphere.
We apply the formulas of the observables in Sec.~IV by using the exact expressions of $\bar{a}$, $\bar{b}$, $\bar{c}$, and $\bar{d}$ 
of the deflection angles in the strong deflection limits without Taylor expansions in the power of an electric charge 
to a supermassive black hole candidate at the center of our galaxy.~\footnote{Note that, from the observation of a black hole shadow 
at the center of the giant elliptical galaxy M87, the Reissner-Nordstr\"{o}m naked singularity there 
is excluded ~\cite{Akiyama:2019eap,Kocherlakota:2021dcv,Akiyama:2019cqa}.}
Our calculations of the observables in Table~1 
are complementary to the case of $q^2/m^2=1.05$ or $q/m\sim 1.025$ investigated by Shaikh~\textit{et al.}~\cite{Shaikh:2019itn}.
As shown Tables~I and II, the total lensed images by the photon sphere around the Reissner-Nordstr\"{o}m naked singularity 
are brighter than the images by the photon sphere around the Reissner-Nordstr\"{o}m black hole by several times
with the exception of an almost marginally unstable photon sphere case.
\footnote{
We have focused on the positive solution of the lens equation 
while there is a negative one $\theta\sim -\theta_{N}$ which makes a pair with the positive one. 
The negative one has almost same magnification as the positive one but its opposite sign. 
Therefore, the total magnification of the pair images is $\mu_{N\mathrm{tot}}=2 \left| \mu_{N} \right|$.
}
Thus, we could distinguish between a Reissner-Nordstr\"{o}m black hole and the naked singularity 
by observations of the images near the photon sphere.

\begin{table*}[htbp]
 \caption{Observables for the images little outside of the photon sphere 
 and the parameters~$\bar{a}$ and $\bar{b}$ in Eq.~(\ref{eq:1}) for given $q/m$.
 We set $D_{\mathrm{OS}}=16$~kpc, $D_{\mathrm{OL}}=D_{\mathrm{LS}}=8$~kpc, and the mass $m= 4\times 10^6 M_{\odot}$.
 The diameter of the photon sphere~$2\theta_{\infty}$ and the outermost image~$2\theta_{\mathrm{E}1}$, 
 the difference of the radii of the outermost image and the photon sphere $\bar{\mathrm{s}}=\theta_{\mathrm{E}1}-\theta_\infty$, 
 the magnification of the pair of the outermost image $\mu_{1\mathrm{tot}}(\phi) \sim 2 \left| \mu_{1} \right|$ 
 for the source angle $\phi=1$ arcsecond, 
 and the ratio of the magnification of the outermost image to the other images $\bar{\mathrm{r}}= \mu_1/\sum^\infty_{N=2} \mu_{N}$
 are shown.   
 }
\begin{center}
\begin{tabular}{ c c c c c c c c c c } \hline\hline
$q/m$          	                          &$0$       &$0.5$     &$1$        &$1.01$    &$1.02$    &$1.03$     &$1.04$   &$1.05$   \\ \hline
$\bar{a}$        	                  &$1.00$    &$1.03$    &$1.41$     &$1.46$    &$1.52$    &$1.61$     &$1.75$   &$2.01$   \\ 
$\bar{b}$        	                  &$-0.400$  &$-0.396$  &$-0.733$   &$-0.821$  &$-0.952$  &$-1.15$    &$-1.53$  &$-2.44$  \\ 
$2\theta_{\infty}$~[$\mu$as]	          &$51.58$   &$49.32$   &$39.71$    &$39.30$   &$38.86$   &$38.38$    &$37.86$  &$37.27$  \\ 
$2\theta_{\mathrm{E}1}$~[$\mu$as]         &$51.65$   &$49.39$   &$40.00$    &$39.60$   &$39.20$   &$38.76$    &$38.29$  &$37.75$  \\ 
$\bar{\mathrm{s}}$~[$\mu$as]              &$0.032$   &$0.038$   &$0.14$     &$0.15$    &$0.17$    &$0.19$     &$0.22$   &$0.24$   \\ 
$\mu_{1\mathrm{tot}}(\phi)\times10^{17}$  &$1.6$     &$1.8$     &$3.8$      &$4.0$     &$4.2$     &$4.4$      &$4.6$    &$4.4$    \\ 
$\bar{\mathrm{r}}$                        &$535$     &$438$     &$85$       &$73$      &$61$      &$49$       &$36$     &$22$     \\ 
\hline\hline
\end{tabular}
\end{center}
\end{table*}

\begin{table*}[htbp]
 \caption{Observables for the images slightly inside of the photon sphere and the parameters $\bar{c}$ and $\bar{d}$ in Eq.~(\ref{eq:2}) for given $q/m$. 
 We assume the same values of $D_{\mathrm{OS}}$ $D_{\mathrm{OL}}$, $m$, $\phi$, and $2\theta_{\infty}$ as Table~I.
 The diameter of the innermost image~$2\theta_{\mathrm{E}1}$, 
 the difference of the radii of the innermost image and the photon sphere $\bar{\mathrm{s}}$, 
 the magnification of the pair of the innermost image $\mu_{1\mathrm{tot}}(\phi)$
 and the ratio of the magnification of the innermost image to the other images $\bar{\mathrm{r}}$
 are shown.   
}
\begin{center}
\begin{tabular}{ c c c c c c } \hline\hline
$q/m$          	                   	       &$1.01$   &$1.02$    &$1.03$    &$1.04$   &$1.05$   \\ \hline
$\bar{c}$    	         	               &$2.92$   &$3.05$    &$3.22$    &$3.49$   &$4.02$   \\ 
$\bar{d}$                	               &$6.01$   &$5.84$    &$5.54$    &$4.97$   &$3.05$   \\ 
$2\theta_{\mathrm{E}1}$~[$\mu$as]              &$28.42$  &$28.47$   &$28.66$   &$29.15$  &$30.43$  \\ 
$\bar{\mathrm{s}}$~[$\mu$as]                   &$-5.44$  &$-5.20$   &$-4.86$   &$-4.36$  &$-3.42$  \\ 
$\mu_{1\mathrm{tot}}(\phi)\times10^{17}$       &$31.9$   &$29.9$    &$27.3$    &$24.0$   &$18.6$   \\ 
$\bar{\mathrm{r}}$                             &$2.5$    &$2.4$     &$2.3$     &$2.1$    &$2.0$    \\ 
\hline\hline
\end{tabular}
\end{center}
\end{table*}

We consider the usual lens configuration only on this paper.
We will investigate retrolensing~\cite{Eiroa:2003jf,Tsukamoto:2016oca,Holz:2002uf,Tsukamoto:2016zdu,Tsukamoto:2017edq}
by the photon sphere in the Reissner-Nordstr\"{o}m naked singularity spacetime on a following paper~\cite{Tsukamoto:2021lpm}.
On this paper, we do not treat the marginally unstable photon sphere for $q/m=3/(2\sqrt{2})$.   
As shown Figure~6, the absolute value of the error of the deflection angle~(\ref{eq:per}) in the strong deflection limit $b\rightarrow b_\mathrm{m}-0$ for 
the winding number $N=1$ violently increases for almost marginally unstable photon sphere case of $q/m\lesssim 3/(2\sqrt{2})$.
In the marginally unstable photon sphere case of $q/m=3/(2\sqrt{2})$, the strong deflection limit analysis totally fails 
because the deflection angle diverges nonlogarithmically. 
In Ref.~\cite{Tsukamoto:2020iez}, Tsukamoto investigated the lensed images barely outside of the marginally unstable photon sphere 
while the inside case is left as future work.
Moreover, naked singularity spacetimes without a photon sphere could make lensed images~\cite{Paul:2020ufc,Dey:2020bgo,Chiba:2017nml}
and more details should be investigated in the future.

\section*{Acknowledgements}
The author thanks Tomohiro Harada for useful discussion and an anonymous referee for valuable comments.
\appendix
\section{Gravitational lensing under a weak-field approximation}
We review shortly gravitational lensing under a weak-field approximation $\left| b \right| \gg  m$. 
In this appendix, both positive and negative impact parameters are considered. 
We can assume $\phi\geq 0$ without loss of generality because of symmetry.
From the deflection angle under the weak-field approximation 
$\alpha \sim 4m/b$,
Eq.~(\ref{eq:10}), and $N=0$,
the solutions of the lens equation~(\ref{eq:11}) is obtained by $\hat{\theta}=\hat{\theta}_{0 \pm}$, where $\hat{\theta}_{0 \pm}$ is defined by  
\begin{eqnarray}
\hat{\theta}_{0 \pm} \left( \hat{\phi} \right) \equiv \frac{1}{2} \left( \hat{\phi} \pm \sqrt{\hat{\phi}^2+4} \right),
\end{eqnarray}
where quantities with the hat denote the quantities divided by an Einstein ring angle $\theta_{\mathrm{E}0}$ which is given by
$\theta_{\mathrm{E}0}\equiv \theta_{0+}(0) \equiv 2\sqrt{mD_{\mathrm{LS}}/(D_{\mathrm{OS}}D_{\mathrm{OL}})}$.
Here and hereinafter, the upper and lower signs are chosen for the positive and negative impact parameters, respectively.
For $D_{\mathrm{OS}}=16$~kpc, $D_{\mathrm{OL}}=D_{\mathrm{LS}}=8$~kpc, and the mass $m= 4\times 10^6 M_{\odot}$,
we get the diameter of the Einstein ring $2\theta_{\mathrm{E}0}=2.86$~arcsecond.
The magnifications of the images are obtained as
\begin{eqnarray}
\mu_{0 \pm}
&\equiv& \frac{\hat{\theta}_{0 \pm}}{\hat{\phi}} \frac{\mathrm{d}\hat{\theta}_{0 \pm}}{\mathrm{d}\hat{\phi}} \nonumber\\
&=& \frac{1}{4} \left( 2\pm \frac{\hat{\phi}}{\sqrt{\hat{\phi}^2+4}} \pm \frac{\sqrt{\hat{\phi}^2+4}}{\hat{\phi}} \right) 
\end{eqnarray}
and the total magnification of the images with the positive and negative impact parameters is given by
\begin{eqnarray}
\mu_{0\mathrm{tot}}
&\equiv& \left| \mu_{0+} \right|+ \left| \mu_{0-} \right| \nonumber\\
&=& \frac{1}{2} \left( \frac{\hat{\phi}}{\sqrt{\hat{\phi}^2+4}} + \frac{\sqrt{\hat{\phi}^2+4}}{\hat{\phi}} \right). 
\end{eqnarray}

\end{document}